\documentclass[conference]{IEEEtran}
\usepackage[a4paper, top=19mm, bottom=43mm, left=14mm, right=14mm]{geometry}

\usepackage{cite}
\usepackage{amsmath,amssymb,amsfonts}
\usepackage{algorithmic}
\usepackage{graphicx}
\usepackage{textcomp}
\usepackage{xcolor}
\def\BibTeX{{\rm B\kern-.05em{\sc i\kern-.025em b}\kern-.08em
    T\kern-.1667em\lower.7ex\hbox{E}\kern-.125emX}}

\usepackage{subcaption}
\usepackage{amsfonts}
\usepackage{bm}
\usepackage{multirow}
\usepackage{booktabs}   
\usepackage{multirow}   
\usepackage{adjustbox}  
\usepackage{enumitem}

\newcommand{\reffig}[1]{Fig. \ref{#1}}

\begin{document}

\title{
  Privacy-Preserving Object Detection for \\
  Vision Transformer-Based Models
\thanks{Identify applicable funding agency here. If none, delete this.}
}

\author{\IEEEauthorblockN{1\textsuperscript{st} Homare Sueyoshi}
\IEEEauthorblockA{\textit{Tokyo Metropolitan University} \\
6-6, Asahigaoka, Hino, Tokyo, Japan \\
sueyoshi-homare@ed.tmu.ac.jp}
\and
\IEEEauthorblockN{2\textsuperscript{nd} Kiyoshi Nishikawa}
\IEEEauthorblockA{\textit{Tokyo Metropolitan University} \\
6-6, Asahigaoka, Hino, Tokyo, Japan \\
kiyoshi@tmu.ac.jp}
\and
\IEEEauthorblockN{3\textsuperscript{rd} Hitoshi Kiya}
\IEEEauthorblockA{\textit{Tokyo Metropolitan University} \\
6-6, Asahigaoka, Hino, Tokyo, Japan \\
kiya@tmu.ac.jp}
}

\maketitle

\begin{abstract}
We propose a novel object detection method that enables us to protect sensitive visual information of test images.
Previous studies considering visual information protection focus on image classification tasks.
This paper proposes an object detection method using perceptual encryption for the first time.
The proposed method can achieve almost the same accuracy as that of models without any protection by utilizing the embedding structure of the Vision Transformer (ViT) and a domain adaptation technique with keys.
In experiments, the effectiveness of the proposed method is verified in terms of accuracy and visual protection under the use of ViTdet, which is a ViT-based object detection model.
\end{abstract}


\section{Introduction}
The vision transformer (ViT) \cite{dosovitskiy2020vit} trained with perceptual encryption has achieved great success in various applications such as privacy-preserving learning \cite{kiya2022overview, hirose2025learnable, sirichotedumrong2018grayscale}, access control, model protection \cite{maungmaung2021protection}, and adversarial defense \cite{aprilpyone2021block, iijima2024random}.
However, previous studies focus on image classification tasks \cite{AprilPyone}.
Accordingly, in this paper, we propose a novel object detection method using perceptional encryption for the first time.
In the proposed method, the ViTdet \cite{li2022vitdet} model, which employs the ViT, is used as a base model for object detection.
Perceptional encryption and a domain adaptation technique with secret keys studied for image classification are expanded for object detection.
In experiments, the proposed method is demonstrated to be effective in object detection in terms of detection accuracy and visual protection.

\section{Related work}

\subsection{ViTDeT}
\begin{figure}[tb]
\includegraphics[width=\linewidth]{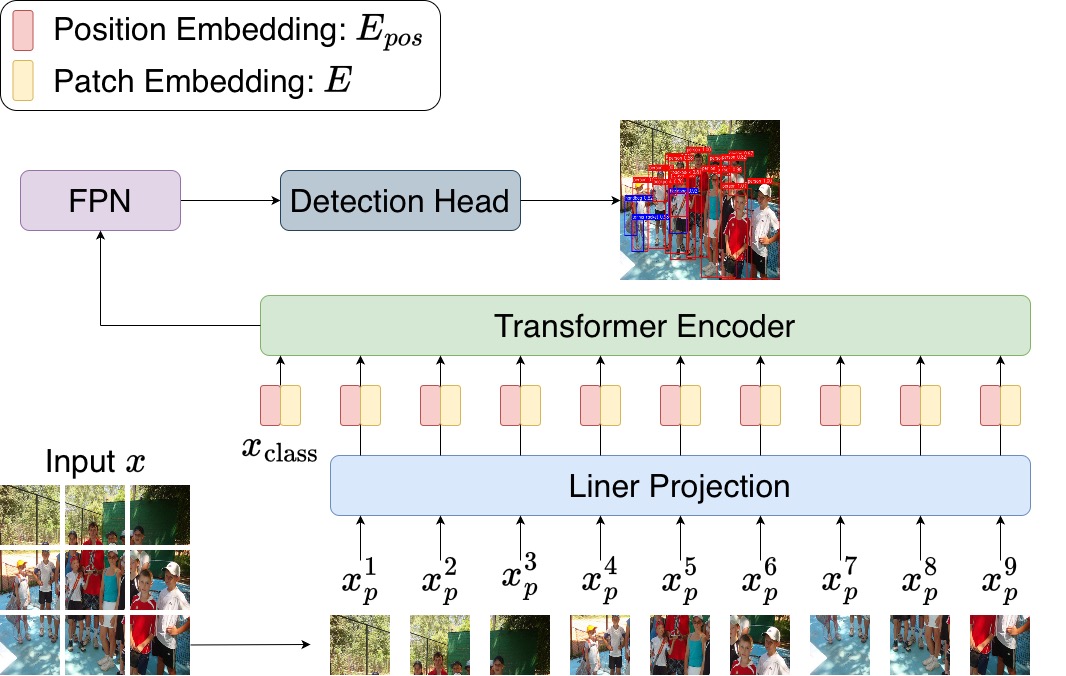}
\caption{Architecture of ViTDeT\label{fig:vitdet_architecture}}
\end{figure}

Our proposed method is designed to be applicable to models based on the Vision Transformer (ViT) architecture.
In this study, we utilize ViTDeT as a representative example of ViT-based Object Detection models.
A architecture of the ViTDeT is shown in Figure~\ref{fig:vitdet_architecture}.
Unlike traditional object detectors that rely on hierarchical backbones and Feature Pyramid Networks (FPNs).
ViTDet adopts a plain ViT backbone.
As shown in the "Input $x$" and "Linear Projection" sections of Figure~\ref{fig:vitdet_architecture}.
Figure~\ref{fig:vitdet_architecture} shows the input image divided into patches $x_p^1, \dots, x_p^9$.
These patches are then linearly projected into an embedding space using a patch embedding $\mathbf{E}$, allowing the model to process image patches similarly to word tokens in natural language processing.
To preserve spatial information, learnable position embeddings $\mathbf{E}_{pos}$ are added to the patch embeddings.
A learnable class token $x_{class}$ is also prepended to the sequence.
Consequently, the input $z_0$ fed into the Transformer Encoder is formulated as follows.
\begin{equation}
    {z}_0= [x_{class}; x_p^1\mathbf{E}; \dots; x_p^N\mathbf{E}] + \mathbf{E}_{pos}
    \label{eq:input_embedding}
\end{equation}
The backbone processes this sequence using Transformer blocks.
While standard ViTs use global attention, ViTDet incorporates Window Attention (computing attention only within local windows) to reduce computational costs for high-resolution inputs, supplemented by a few Global Attention blocks to enable information propagation across windows.
Finally, a simple feature pyramid is constructed from the single-scale feature map output to perform object detection.

In this paper, specifically within the structure described in Equation~\ref{eq:input_embedding}, we focus on the patch embedding matrix $\mathbf{E}$.

\section{Proposed Method}
\subsection{Overview of propoesd method}
\begin{figure}[tb]
  \centering
  \includegraphics[width=\linewidth]{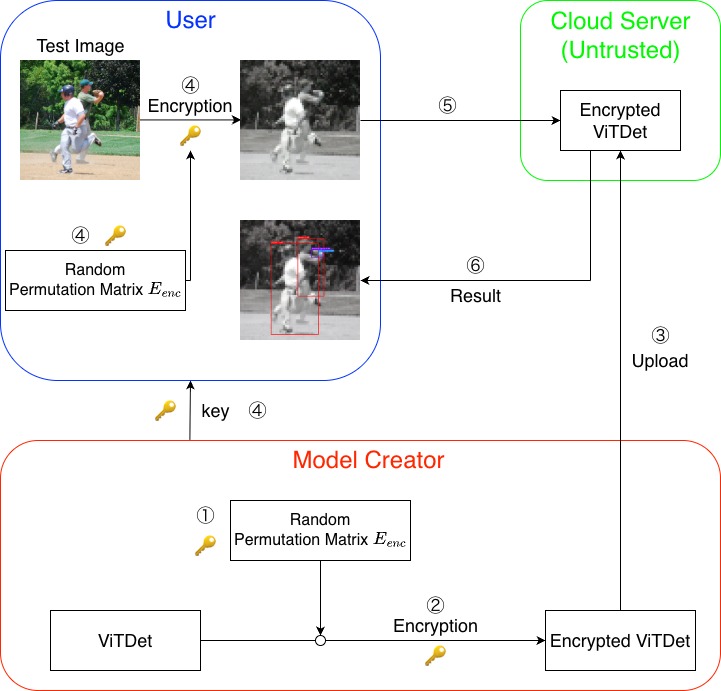}
  \caption{Framework of proposed method}
  \label{fig:proposed_method}
\end{figure}

The framework of the proposed method is summarized as below (see \reffig{fig:proposed_method}).

\begin{enumerate}[label=\textcircled{\raisebox{-0.7pt}{\small\arabic*}}]
  \item \label{step:encrypt}
	A model creator generates a random permutation matrix $\bm{E}_{\mathrm{enc}}$ using a secret key $k$.

	\item \label{step:encrypt2}
  Encrypt patch embedding $\bm{E}$ to get an encrypted model by using $\bm{E}_{\mathrm{enc}}$ in a model trained with plain images.

  \item
  Upload the encrypted model to a cloud server.

  \item
  A user receives $\bm{E}_{\mathrm{enc}}$ from the model creator and encrypt a test image using $\bm{E}_{\mathrm{enc}}$.

  \item
  Upload the encrypted test image to the cloud server.

  \item
  The client receives an inference result from the cloud server.
\end{enumerate}

In the above framework, the cloud server does not have plain images and the secret key.

\subsection{Generation of Random Permutation Matrix}
\label{sec:random_substitution_matrix}

The generation process of the random permutation matrix is described as follows:

\begin{enumerate}[label=\arabic*), ref=\arabic*)]
  \item
  A vector is given by:
	\begin{equation}
    \bm{l} = [l(1), l(2), \dots, l(i), \dots, l(L)],
  \end{equation}
  where each element $l(i)$ satisfies:
  \begin{gather*}
    l(i) \in \{1, 2, \dots, L\} \\
    i \neq j \Rightarrow l(i) \neq l(j), \quad \forall i, j \in \{1, 2, \dots, L\}.
  \end{gather*}
	$L = p^{2}c$ is the number of pixels in a block.
  $p$ is defined as the patch size, and $c$ is defined as the number of channels in the image.

  \item
  A random permutation matrix $\bm{E}_{\mathrm{enc}} \in \mathbb{R}^{L \times L}$ is defined as:
  \begin{equation}
    \bm{E}_{\mathrm{enc}} =
    \begin{bmatrix}
      k_{(1,1)} & k_{(1,2)} & \cdots & k_{(1,L)} \\
      k_{(2,1)} & k_{(2,2)} & \cdots & k_{(2,L)} \\
      \vdots & \vdots & \ddots & \vdots \\
      k_{(L,1)} & k_{(L,2)} & \cdots & k_{(L,L)}
    \end{bmatrix}
    .
  \end{equation}
  where the components $k_{(i,j)}$ are determined by:
  \begin{gather*}
    k_{(i,j)} =
    \begin{cases}
      0 & \text{if } l(i) \neq j \\
      1 & \text{if } l(i) = j
    \end{cases}
    .
  \end{gather*}
  Since $\bm{E}_{\mathrm{enc}}$ is an orthogonal matrix, the following property holds:
  \begin{equation}
    \bm{E}_{\mathrm{enc}}^{-1} = \bm{E}_{\mathrm{enc}}^{\top}.
  \end{equation}
\end{enumerate}

As a concrete example, if $L=4$ and the generated vector is $\bm{l}=[3, 4, 2, 1]$,
the random permutation matrix $\bm{E}_{\mathrm{enc}}$ is given by:
\begin{equation}
  \bm{E}_{\mathrm{enc}} =
  \begin{bmatrix}
    0 & 0 & 1 & 0 \\
    0 & 0 & 0 & 1 \\
    0 & 1 & 0 & 0 \\
    1 & 0 & 0 & 0
  \end{bmatrix}
  .
\end{equation}

\subsection{Image Encryption}
\begin{figure}[tb]
\includegraphics[width=\linewidth]{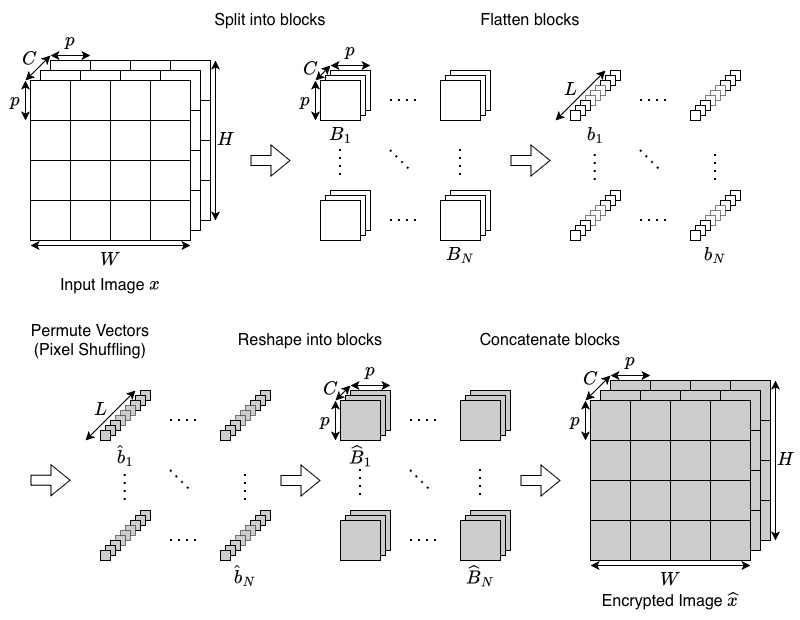}
\caption{Procedure of image encryption\label{fig:image_enc_procedure}}
\end{figure}

The image encryption procedure is shown in the Figure~\ref{fig:image_enc_procedure}.
Encrypted images are used for model testing.
In the proposed method, an encryption method is carried out in a similar manner as in conventional methods \cite{jimaging8090233}.
An input image $\bm{x}$ is divided into $N$ non-overlapping blocks $\bm{B} = \{B_{1}, B_{2}, \dots, B_{N}\}$ of size $p \times p$.
$N = \frac{W}{p} \times \frac{H}{p}$ is the number of blocks.
We assume that width ($W$) and height ($H$) are divisible by $p$.
The block size $p$ is set to be consistent with the patch size $p$ of the ViT.
Each block $\bm{B}_i$ is flattened into a vector $\bm{b}_i$.
Using $\bm{E}_{\mathrm{enc}} \in \mathbb{R}^{L \times L}$, each image block is encrypted as follows:
\begin{align}
  \hat{\bm{b}}_i &= \bm{b}_i \bm{E}_{\mathrm{enc}}^{\top}.
\end{align}
The encrypted vectors $\hat{\bm{b}}_i$ are subsequently reshaped into encrypted blocks $\hat{\bm{B}}_i$, which are then concatenated to construct the encrypted image $\hat{\bm{x}}$.

\begin{figure}[tb]
  \centering
  \captionsetup[subfigure]{labelformat=empty}
  \begin{minipage}[b]{0.48\linewidth}
    \centering
    \includegraphics[width=0.95\textwidth]{img/coco_plain.jpg}
    \subcaption{Original image}
    \label{fig:cityscapes_plain}
  \end{minipage}
  \begin{minipage}[b]{0.48\linewidth}
    \centering
    \includegraphics[width=0.95\textwidth]{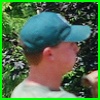}
    \subcaption{Zoomed-in view}
    \label{fig:cityscapes_plain_crop}
  \end{minipage}
  \caption*{(a) Non-encrypted images}

  \vspace{0.5em}

  \begin{minipage}[b]{0.48\linewidth}
    \centering
    \includegraphics[width=0.95\textwidth]{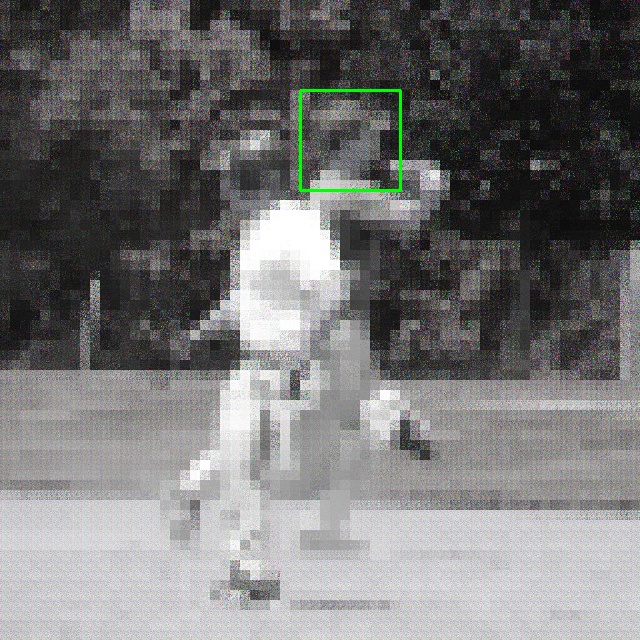}
    \subcaption{Encrypted image}
    \label{fig:cityscapes_enc}
  \end{minipage}
  \begin{minipage}[b]{0.48\linewidth}
    \centering
    \includegraphics[width=0.95\textwidth]{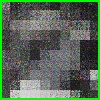}
    \subcaption{Zoomed-in view}
    \label{fig:cityscapes_enc_crop}
  \end{minipage}
  \caption*{(b) Encrypted images (block size: 16×16)}

  \caption{Example of encrypted images (image size: 1024×1024)}
  \label{fig:cityscapes_comparison}
\end{figure}

\reffig{fig:cityscapes_comparison} shows an example of encrypted images.

\subsection{Model Encryption}
The trained model is encrypted by using the equation
\begin{equation}
    \hat{\mathbf{E}} = \mathbf{E}_{enc}\mathbf{E}.
\end{equation}
where $\bm{E} \in \mathbb{R}^{D \times L}$ and $D$ denotes the embedding dimension of the Vision Transformer model
(see \reffig{fig:proposed_method}).
Accrordingly, the inference is carried out in the encrypted domain.

\subsection{Properties of proposed method}
We analyze the impact of encryption on object detection accuracy.
Let $\mathbf{E}$ be the original patch embedding projection matrix and $x_p^i$ be the $i$-th patch of the input image.
We introduce an encryption matrix $\mathbf{E}_{enc}$, which satisfies the orthogonality condition $\mathbf{E}_{enc}^\top = \mathbf{E}_{enc}^{-1}$.

The encrypted embedding matrix $\hat{\mathbf{E}}$ and the encrypted $i$-th patch $\widehat{x_p^i}$ are defined as follows:
\begin{equation}
    \hat{\mathbf{E}} = \mathbf{E}_{enc}\mathbf{E}, \quad \widehat{x_p^i} = x_p^i \mathbf{E}_{enc}^\top.
\end{equation}
By substituting these definitions, the projection of the encrypted patch can be derived as:
\begin{equation}
    \widehat{x_p^i}\hat{\mathbf{E}} = (x_p^i \mathbf{E}_{enc}^\top)(\mathbf{E}_{enc}\mathbf{E}) = x_p^i (\mathbf{E}_{enc}^\top \mathbf{E}_{enc}) \mathbf{E} = x_p^i \mathbf{E},
\end{equation}
where we utilize the property that $\mathbf{E}_{enc}^\top \mathbf{E}_{enc} = \mathbf{I}$.

Consequently, the input sequence $\hat{z}_0$ for the encrypted model is given by:
\begin{equation}
\begin{aligned}
    \hat{z}_0 &= [x_{class}; \widehat{x_p^1}\hat{\mathbf{E}}; \dots; \widehat{x_p^N}\hat{\mathbf{E}}] + \mathbf{E}_{pos} \\
    &= [x_{class}; x_p^1\mathbf{E}; \dots; x_p^N\mathbf{E}] + \mathbf{E}_{pos} \\
    &= z_0
\end{aligned}
\end{equation}
This result demonstrates that $ \hat{z}_0 = z_0 $, meaning that the model encryption effectively cancels out the influence of the image encryption, allowing the Transformer to process the data without any influence of encryption on object detection accuracy.




\section{Experiments}
\subsection{Setup}
In experiments, an object detection task was performed by using the COCO dataset \cite{lin2015coco} as test images, which consists of 80 object categories.
The performance of trained models was also evaluated with  the ``val2017'' subset as validation images from the COCO dataset.

Furthermore, we extended our evaluation to the LVIS dataset \cite{gupta2019lvis}.
While LVIS shares the exact same images as those of the COCO dataset, it is characterized by different data splits and significantly richer annotations.
Unlike the 80 categories in COCO, LVIS features a massive vocabulary of 1,203 categories.
To comprehensively assess detection performance, we adopted ViTdet as a vision transformer-based model with the Mask R-CNN framework.


For evaluation metrics, we employed the standard Average Precision (AP) scores, which range from 0 to 100, where a higher value indicates superior performance.
To evaluate performance across different object scales, we utilized $\text{AP}_{s}$, $\text{AP}_{m}$, and $\text{AP}_{l}$ for small (area $< 32 \times 32$), medium ($32 \times 32 \le \text{area} \le 96 \times 96$), and large (area $> 96 \times 96$) objects.

\subsection{Experimental Results}
\label{subsec:experiment_result}
Table~\ref{tab:mask_rcnn_coco} presents the evaluation results of object detection accuracy (AP) under various conditions.
In this evaluation, ViTdet models with three different backbones (ViT-B, ViT-L, and ViT-H) were utilized.
These models correspond to the three models called ``model\_final\_61ccd1.pkl'', ``model\_final\_6146ed.pkl'', and ``model\_final\_7224f1.pkl'' published on the official ViTdet GitHub repository.
``Baseline'' represents the results obtained using both non-encrypted models and images.
``Without model encryption'' shows results when only the images were encrypted and the models were not encrypted.
In the table, ``Proposed'' denotes the proposed method utilizing both encrypted model and images.
As shown in Table~\ref{tab:mask_rcnn_coco}, it was confirmed that the proposed method achieved almost the same as that of the non-encrypted models.
Theoretically, the proposed block-wise encryption can cancel out the influence of the image encryption at the embedding layer, but the accuracy was slightly different from that of ``Baseline'' due to the influence of JPEG compression used in the COCO dataset.

\begin{table}[tb]
\caption{Comparison of object detection accuracy (AP) using Mask R-CNN on COCO dataset.\label{tab:mask_rcnn_coco}}
\centering
\begin{adjustbox}{width=\columnwidth}
\begin{tabular}{cccccccc}
\toprule
\textbf{Backbone} & \textbf{Method} & \textbf{AP} & \textbf{AP$_{s}$} & \textbf{AP$_{m}$} & \textbf{AP$_{l}$} \\
\midrule
\multirow{3}{*}{ViT-B} & Baseline                 & 51.412 & 35.645 & 55.031 & 66.273 \\
                       & Without model encryption & 0.035  & 0.018  & 0.055  & 0.156  \\
                       & Proposed                 & 50.118 & 32.882 & 53.875 & 64.717 \\
\midrule
\multirow{3}{*}{ViT-L} & Baseline     & 55.450 & 39.342 & 59.699 & 69.569 \\
                       & Without model encryption & 0.804  & 0.129  & 0.648  & 1.776  \\
                       & Proposed     & 54.534 & 38.762 & 58.634 & 68.337 \\
\midrule
\multirow{3}{*}{ViT-H} & Baseline     & 56.594 & 40.456 & 60.783 & 71.301 \\
                       & Without model encryption & 0.406  & 0.066  & 0.733  & 0.725  \\
                       & Proposed     & 55.843 & 40.895 & 59.932 & 69.997 \\
\bottomrule
\end{tabular}
\end{adjustbox}
\end{table}

\begin{table}[t]
\caption{Comparison of object detection accuracy (AP) using Mask R-CNN on LVIS dataset.\label{tab:mask_rcnn_lvis}}
\centering
\begin{adjustbox}{width=\columnwidth}
\begin{tabular}{ccccccccccc}
\toprule
\textbf{Backbone} & \textbf{Method} & \textbf{AP} & \textbf{AP$_{s}$} & \textbf{AP$_{m}$} & \textbf{AP$_{l}$} \\
\midrule
\multirow{3}{*}{ViT-B} & Baseline                 & 39.883 & 30.731 & 48.922 & 56.971 \\
                       & Without model encryption & 0.030  & 0.009  & 0.026  & 0.080  \\
                       & Proposed                 & 38.585 & 29.339 & 47.601 & 55.375 \\
\midrule
\multirow{3}{*}{ViT-L} & Baseline                 & 45.924 & 36.345 & 55.264 & 63.392 \\
                       & Without model encryption & 0.286  & 0.048  & 0.284  & 1.002  \\
                       & Proposed                 & 44.546 & 35.738 & 53.989 & 61.469 \\
\midrule
\multirow{3}{*}{ViT-H} & Baseline                 & 49.123 & 39.369 & 59.184 & 65.992 \\
                       & Without model encryption & 0.227  & 0.070  & 0.230  & 0.611  \\
                       & Proposed                 & 47.811 & 38.917 & 57.320 & 64.993 \\
\bottomrule
\end{tabular}
\end{adjustbox}
\end{table}

\begin{table}[t]
\caption{Comparison of object detection accuracy (AP) using Cascade Mask R-CNN on COCO dataset.\label{tab:cascade_mask_rcnn_coco}}
\centering
\begin{adjustbox}{width=\columnwidth}
\begin{tabular}{cccccccc}
\toprule
\textbf{Backbone} & \textbf{Method} & \textbf{AP} & \textbf{AP$_{s}$} & \textbf{AP$_{m}$} & \textbf{AP$_{l}$} \\
\midrule
\multirow{3}{*}{ViT-B} & Baseline                 & 53.781 & 36.146 & 57.487 & 68.660 \\
                       & Without model encryption & 0.085  & 0.013  & 0.116  & 0.335 \\
                       & Proposed                 & 52.520 & 34.307 & 56.028 & 67.181 \\
\midrule
\multirow{3}{*}{ViT-L} & Baseline                 & 57.470 & 40.699 & 61.592 & 72.414 \\
                       & Without model encryption & 0.739  & 0.047  & 0.525  & 1.704  \\
                       & Proposed                 & 56.671 & 40.326 & 60.651 & 71.172 \\
\midrule
\multirow{3}{*}{ViT-H} & Baseline                 & 58.650 & 41.985 & 62.684 & 73.835 \\
                       & Without model encryption & 0.399  & 0.071  & 0.484  & 0.603  \\
                       & Proposed                 & 57.586 & 41.347 & 61.868 & 72.262 \\
\bottomrule
\end{tabular}
\end{adjustbox}
\end{table}

\begin{table}[tb]
\caption{Comparison of object detection accuracy (AP) using Cascade Mask R-CNN on LVIS dataset.\label{tab:cascade_mask_rcnn_lvis}}
\centering
\begin{adjustbox}{width=\columnwidth}
\begin{tabular}{ccccccccccc}
\toprule
\textbf{Backbone} & \textbf{Method} & \textbf{AP} & \textbf{AP$_{s}$} & \textbf{AP$_{m}$} & \textbf{AP$_{l}$} \\
\midrule
\multirow{3}{*}{ViT-B} & Baseline                 & 42.770 & 32.840 & 52.097 & 61.230 \\
                       & Without model encryption & 0.074  & 0.005  & 0.027  & 0.260  \\
                       & Proposed                 & 41.570 & 31.500 & 51.168 & 59.783 \\
\midrule
\multirow{3}{*}{ViT-L} & Baseline                 & 49.017 & 38.651 & 59.417 & 67.392 \\
                       & Without model encryption & 0.214  & 0.024  & 0.133  & 0.732  \\
                       & Proposed                 & 47.900 & 37.950 & 57.720 & 65.849 \\
\midrule
\multirow{3}{*}{ViT-H} & Baseline                 & 51.175 & 41.012 & 61.324 & 68.627 \\
                       & Without model encryption & 0.131  & 0.027  & 0.159  & 0.383  \\
                       & Proposed                 & 50.583 & 40.755 & 60.442 & 67.860 \\
\bottomrule
\end{tabular}
\end{adjustbox}
\end{table}

\section{Conclusion}
We proposed a privacy-preserving method for object detection that utilizes the embedding structure of Vision Transformers (ViTs).
Using a ViTdet as a representative example, we demonstrated that applying block-wise encryption effectively protects visual information while maintaining high detection accuracy equivalent to that of non-encrypted models.
Furthermore, the significant decrease in accuracy observed when only the images were encrypted demonstrates the necessity of encrypting the model's embedding layer.

\section*{ACKNOWLEDGEMENTS}
This work was supported in part by JSPS KAKENHI (Grant Number 25K07750).

\end{document}